\documentclass[sigconf]{acmart}
\usepackage{appendix}
\usepackage{enumitem} 
\usepackage[utf8]{inputenc}
\usepackage{subcaption}

\AtBeginDocument{%
  }

\copyrightyear{2026}
\acmYear{2026}
\setcopyright{cc}
\setcctype{by}
\acmConference[ETRA '26]{2026 Symposium on Eye Tracking Research and Applications}{June 01--04, 2026}{Marrakesh, Morocco}
\acmBooktitle{2026 Symposium on Eye Tracking Research and Applications (ETRA '26), June 01--04, 2026, Marrakesh, Morocco}
\acmDOI{10.1145/3797246.3803036}
\acmISBN{979-8-4007-2519-7/2026/06}

\acmSubmissionID{1074}

\begin{document}



\title{Understanding Password Preferences, Memorability, and Security through a Human-Centered Lens}


\author{Duru Paker}
\email{duru.paker@tum.de}
\affiliation{%
  \institution{Technical University of Munich}
  \city{Munich}
  \country{Germany}
}

\author{Suleyman Ozdel}
\email{ozdelsuleyman@tum.de}
\affiliation{%
  \institution{Technical University of Munich, Munich Center for Machine Learning}
  \city{Munich}
  \country{Germany}
}

\author{Enkelejda Kasneci}
\email{enkelejda.kasneci@tum.de}
\affiliation{%
  \institution{Technical University of Munich, Munich Center for Machine Learning}
  \city{Munich}
  \country{Germany}
}


\begin{abstract}

Passwords remain the primary authentication method, yet user-created passwords are often the weakest due to the security–usability trade-off. Although AI-based password generators are emerging, little is known about their effectiveness and user perceptions. This eye-tracking study examined how behavior during password creation, selection, and memorization relates to objective and subjective password quality. Four password models, three AI-based (DeepSeek-API, ChatGPT-API, PassGPT) and one rule-based random generator, generated suggestions from participants’ self-generated passwords across four website contexts. Eye movements were recorded throughout the experiment. Results confirm the expected trade-off between AI-generated password strength and human memorability but also reveal a novel behavioral link. Despite stronger AI-generated passwords, participants favored self-generated ones. Notably, visual attention to contextual cues was significantly correlated with higher password entropy. This suggests that security is shaped not only by the generation tool but also by users’ visual engagement with contextual cues, highlighting the potential of attention-driven security design.

\end{abstract}



\begin{teaserfigure}
  \centering
  \includegraphics[width=0.7\textwidth]{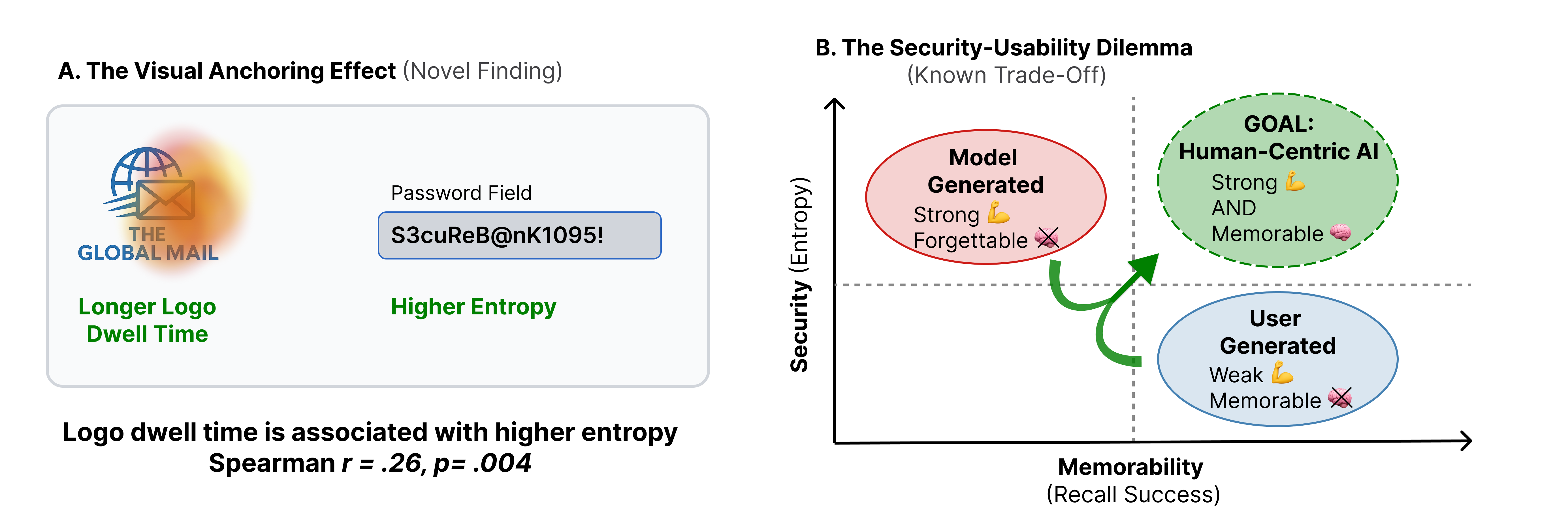} 
  \caption{\textbf{The Visual Anchoring Effect and the Security-Usability Trade-off.} (A) Our novel eye-tracking finding reveals that increased visual attention (dwell time) to contextual cues like service logos is significantly associated with the creation of higher-entropy passwords ($N=15$ participants, 120 trials). (B) A conceptual map of the known trade-off: model-generated passwords are strong but forgettable, while user-generated ones are memorable but weak. We propose "visual anchoring" as a design intervention to nudge users toward the human-centric goal of high security and high memorability.}
  \Description{A two-panel teaser figure. Panel A on the left, titled 'The Visual Anchoring Effect (Novel Finding)', shows a login screen schematic with a bank shield logo. A glowing orange heatmap blob is over the logo, labeled 'Gaze Heatmap (Dwell Time)'. An arrow points from the logo to a password field containing 'S3cur3#B@nk!99' with a small 'Higher Entropy' label next to it. Text below reads: 'Logo dwell time is associated with higher entropy'. Below that, large bold statistics read: 'Spearman r = .26, p = .004'. A note at the bottom says 'N = 15 participants (120 trials)'. Panel B on the right, titled 'The Security-Usability Dilemma (Known Trade-off)', is a 2D conceptual chart. The Y-axis is 'SECURITY (Entropy)' and the X-axis is 'MEMORABILITY'. A red circle in the top-left is labeled 'Model-Generated: Strong but Forgettable'. A blue dashed circle in the bottom-right is labeled 'User-Generated: Familiar but Weak'. A curved arrow labeled 'visual anchoring nudge' points from the blue circle to a dark dotted ellipse in the top-right labeled 'GOAL: Human-Centric AI (The 'Sweet Spot') high security + high memorability'. A tiny note at the bottom reads 'Conceptual schematic (not to scale)'.}
  \label{fig:teaser}
\end{teaserfigure}


\maketitle

\section{Introduction}

As technology increasingly shapes and transforms every aspect of our lives, such as shopping and even banking, passwords remain the dominant authentication mechanism, despite the availability of more secure alternatives. This gives users a central role in securing their profiles, although user-generated passwords are often the weakest link \cite{Alavanza24}. Balancing security and usability remains challenging for users, which leads to weak or reused passwords \cite{Shay10}. In recent years, Artificial Intelligence (AI) driven password generators and Large Language Model (LLM) based suggestions have emerged as tools to help users balance security and usability.

Prior studies have examined both technical and behavioral aspects of password security, including password strength metrics, policy effects on user behavior \cite{Ur2012-1,Weir10,Shay16}, and the use of gaze behavior and pupil dilation as predictors of password strength \cite{Abdrabou21_3,Abdrabou212}. However, these perspectives have rarely been jointly examined from a human-centered perspective in the context of AI-based password suggestion systems.

This study examines user behavior during password creation, selection, and memorization across different password-generation models and contexts. Specifically, we investigate (RQ1) how different password generation methods differ in terms of password strength, memorability, and user perception; (RQ2) how contextual factors shape password behavior and gaze patterns; and (RQ3) how gaze-based indicators, subjective perception, and behavioral measures relate to objective password quality and memorization performance.

To address these questions, an experiment was conducted using a Tobii Pro Fusion eye tracker. Participants created passwords for four contexts and chose either their own password or one suggestion from four models: three AI-based and one rule-based (PassGPT, GPT-API, DeepSeek-API, and a random generator). Participants subsequently evaluated their chosen passwords and completed a memorability task while gaze and behavioral data were collected.

In this work, we investigate how users interact with AI-generated and self-generated passwords during creation and memorization, and how gaze behavior relates to password quality and memorability was investigated. Eye tracking was used to record users’ visual attention and eye-movement dynamics while they create, select, and recognize passwords generated by different models and across different website contexts. Results show a clear trade-off between password strength and memorability, and reveal that users’ perceptions of password strength do not fully reflect objective security differences. We further find that increased visual attention to contextual cues, such as service logos, is associated with higher password entropy. To the best of our knowledge, this is the first study to combine eye-tracking measures with AI-assisted password generation to examine how cognitive effort, user perception, and password quality interact. These findings provide insight for the design of human-centered AI-assisted password systems that better balance security and usability.


\section{Related Work}

Human-chosen passwords remain the most widely used authentication mechanism, yet weak passwords continue to undermine security \cite{Alavanza24}. Numerous password-cracking techniques have been developed \cite{Najafabadi14, Cubrilovic09, DeCarne14}. As a result, providing users with guidance to generate stronger passwords has become common practice \cite{Ur2012-1}.

Although strong passwords mitigate attacks, users often prioritize memorability over security. Password policies can steer users toward stronger passwords \cite{Proctor2002}. However, overly strict requirements increase frustration and reuse, allowing security gaps \cite{Inglesant10, Shay10, Zhang10}. Prior work therefore emphasizes balancing security and usability, for example by emphasizing password length over complex character requirements \cite{Weir10, Ur2012-2, Shay16}. Novel approaches to capture complex structural dependencies focused on data-driven password generation \cite{Cui20, Dell'Amico15, Thai24}. This shift has laid the foundation for modern deep learning-based and large language model-based password generation approaches. 

\subsection{Deep Learning and LLM-Based Password Generation}

To overcome the limitations of rule-based and heuristic password guessing and generation methods, prior work has applied deep learning models that learn password distributions from large-scale leaked credential datasets \cite{Zhang20, Pal19}. Neural Networks outperform other individual deep-learning-based password-generation methods when targeting more complex passwords and at a higher number of guesses \cite{Melicher16}.

Large Language Models based on transformer architectures demonstrate strong complex sequence-modeling capabilities, making them suitable for password generation. By modeling passwords at the character level, PassGPT, an LLM-based model \cite{Rando23}, outperformed prior approaches such as PassGAN \cite{Hitaj19}. In 2024, PagPassGPT built upon PassGPT by introducing a divide-and-conquer approach to reduce repetitions, thereby improving password diversity and quality. PagPassGPT achieves up to a 27.5\% higher hit rate than PassGPT \cite{Su24}.

\subsection{Eye Tracking and Passwords}

Eye tracking provides insights into cognitive load and attention allocation during interaction. Metrics such as fixation duration, saccades, and pupil dilation have been widely used as indicators of mental workload~\cite{Liu22, Chen11, Chien15, Goldberg99}. Pupil size is also a strong indicator of cognitive activity \cite{Chen14,Gavas17}. Together, these measures can help identify how much effort a person is investing. 

Several studies have shown that eye tracking can provide useful information about password creation and security. Pupil size was examined while people created weak and strong passwords \cite{Abdrabou21_3}. Larger pupils pointing towards higher cognitive load were observed during strong password creation. In a follow-up study, fixations, saccades, and pupil measures during password creation were analyzed \cite{Abdrabou212}. From this, weak vs. strong passwords could be classified with up to 86\% accuracy. Furthermore, another study predicted password reuse from gaze and typing behavior \cite{Abdrabou22}. Almost 89\% success in predicting password reuse was reached.


\paragraph{Gap and Motivation:}While prior work has examined AI-based password generation and gaze-based indicators of password behavior, these study topics were mostly separate. Existing eye-tracking studies focus primarily on user-generated passwords, whereas evaluations of LLM-based password generators emphasize strength and guessability, with limited attention to user interaction, perception, or memorability. As a result, it is still unclear how users visually and cognitively engage with AI-assisted password suggestions, or how visual attention relates to both objective password quality and memorability. Addressing this gap is necessary for assessing the future of AI-assisted password generation.

\section{Methodology}

\subsection{Participants and Apparatus}
Fifteen participants (mean age= 24.6, range = 22-32, 9 male, 6 female) were recruited from university. Most participants were students (n=11), and all reported regular use of AI tools. The experiment was implemented as a web-based survey using the Flask framework and completed on a desktop computer with a screen resolution of 1920$\times$1080 pixels in a computer laboratory via a browser. The interface was implemented using HTML and used the browser’s default font settings (approximately 16\,px). Throughout the experiment, on-screen coordinates of interface elements were logged, structured into Areas of Interest (AOIs). Password suggestions were generated using PassGPT, GPT-API, DeepSeek-API, and a rule-based baseline. All password suggestions, user-generated passwords, and memorized passwords were recorded in anonymized form. Eye movements were recorded using the screen-based Tobii Pro Fusion eye tracker at 250 Hz. Raw gaze data included gaze points and pupil diameter, and data collection was controlled via the Tobii Pro SDK integrated into the experiment. A standard five-point calibration was performed before the experiment and repeated after the break.

\subsection{Experimental Design and Procedure}

\begin{figure}[htbp]
    \centering  
    \begin{subfigure}[b]{0.4\textwidth}
        \caption{Logged Coordinates on "Choose" Screen} 
        \includegraphics[width=\linewidth]{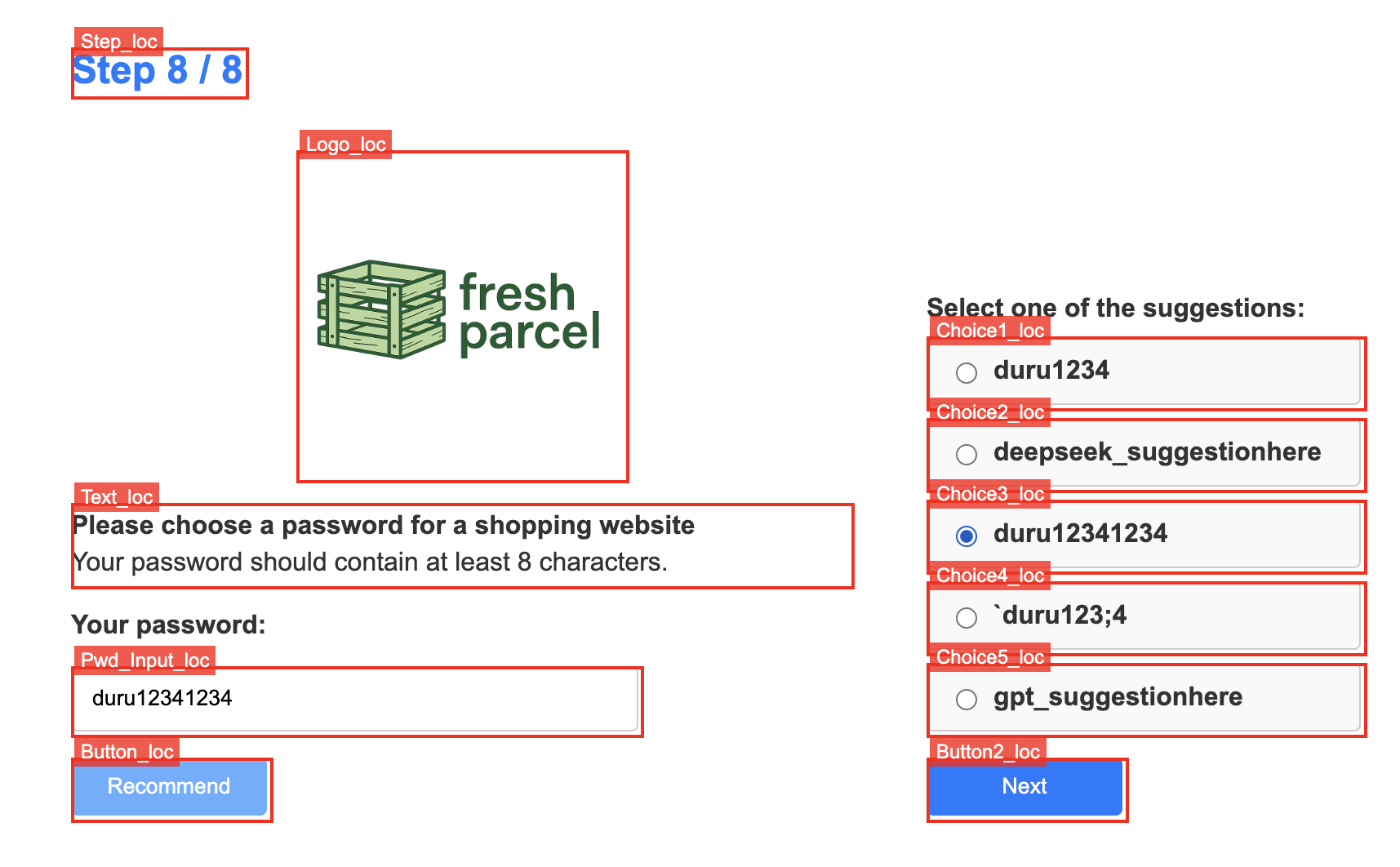}
        \label{fig:choose}
    \end{subfigure}
    \hspace{0.1\textwidth}
    \begin{subfigure}[b]{0.2\textwidth}
        \caption{Logged Coordinates on "Memorability" Screen} 
        \includegraphics[width=\linewidth]{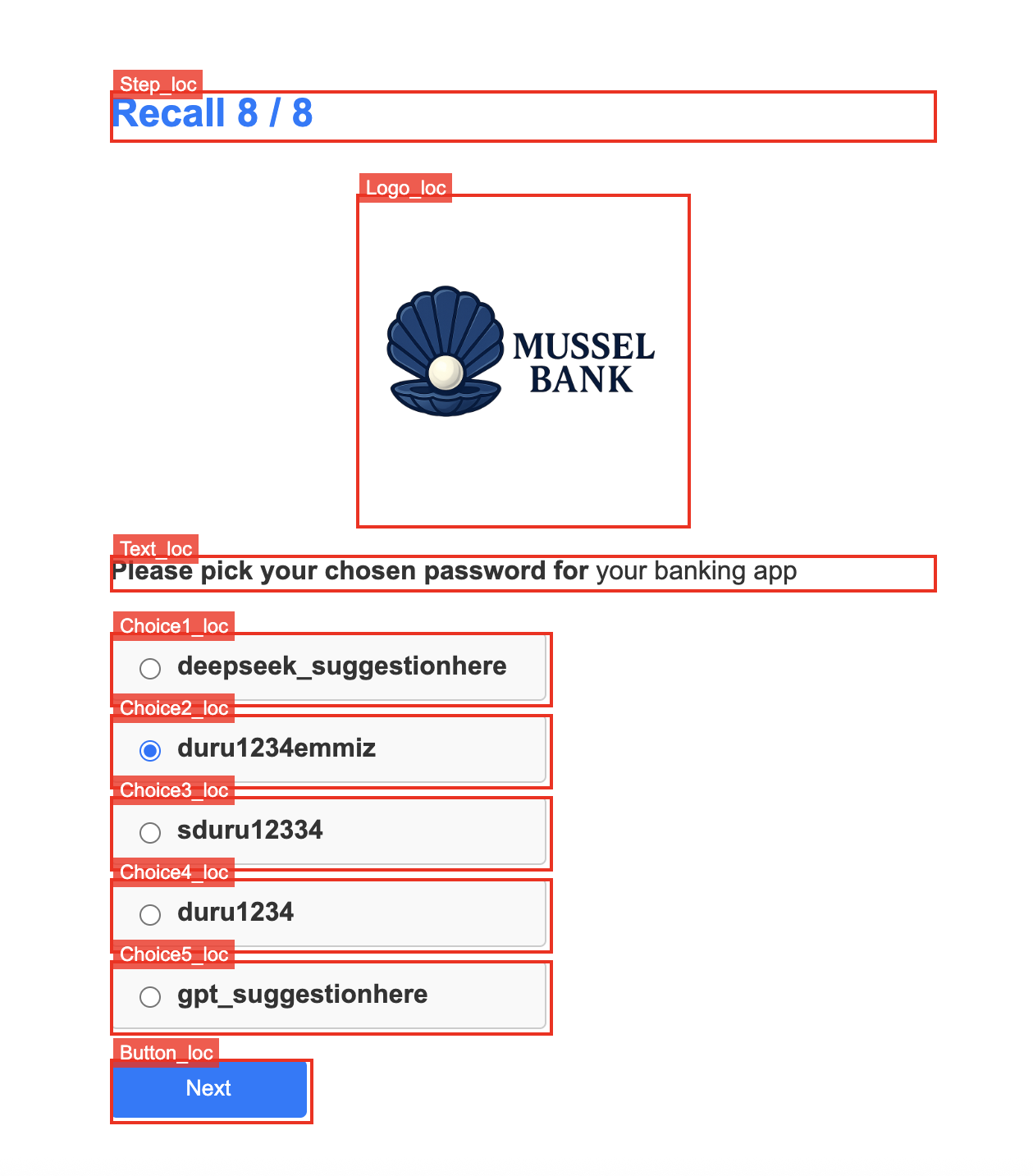}
        \label{fig:recall}
    \end{subfigure}
    \caption{Examples of AOI logging, red boxes were not visible to participants during the experiment.}
    \Description{Two screenshots of the experiment interface. The left panel shows the password selection screen with five password options, each highlighted with a red bounding box indicating the logged AOI coordinates. The right panel shows the memorability assessment screen with five password options similarly annotated with red AOI boxes.}
    \label{coordinates}
\end{figure}

The experiment lasted approximately 20–25 minutes and consisted of three phases: password creation and selection, memorability assessment, and evaluation. Participants completed tasks for eight fictional websites spanning four contexts (banking, shopping, email, social media), presented in randomized order. After confirming the consent form and completing the eye-tracker calibration, participants completed the "input and choose" phase.

During password creation, participants were instructed to enter a password of at least eight characters, consistent with NIST standards \cite{Grassi2017}. In the choose section, participants were shown five password options and required to select one before proceeding. Suggestions were displayed in identical format, with randomized order and no source information shown. Figure \ref{fig:choose} illustrates the logged on-screen coordinates and defined AOIs for the "Choose" screen. The five options consisted of the participant’s own password and four model-generated suggestions. One suggestion was generated by PassGPT (passgpt-16characters) ~\cite{Rando2316}, a GPT-2–based model ~\cite{Rando23}. Additional suggestions were generated using OpenAI’s GPT-4o model ~\cite{openai_gpt4o_2024} and the DeepSeek-API (deepseek-chat) ~\cite{deepseek_chat_2025}, both using the same prompt defining the model as a “Password Booster” (see Appendix ~\ref{app:prompt_gpt}). As a baseline, a Persuasive Text Password generator inserted 2–4 random characters into random positions of the original password ~\cite{Nguyen15}.

After completing all eight input-and-choose tasks, participants took a five-minute break and were instructed to keep their chosen passwords in mind. Calibration was repeated shortly after the break. In the memorability phase, participants selected the previously chosen password from five options, with website order randomized again. As the study focuses on user preferences and usability of AI-generated passwords, memorability was assessed for relative comparison between models and user-generated passwords. Therefore, we used a five-option multiple-choice recognition task rather than free recall to reduce sensitivity to minor recall errors (e.g., single-character mistakes) and to better capture practical memorability. Figure \ref{fig:recall} illustrates the "Memorability" screen. Subsequently, in the evaluation phase, participants rated the memorability and strength of their selected passwords using five-point Likert scales. Finally, participants completed a demographic survey.

\subsection{Measurements}

\subsubsection{Eye Tracking Measures}

The collected raw eye-tracking data included device and system timestamps, pupil diameter for both eyes, gaze points on the display area and gaze origin. From these raw data, gaze-based measures were derived as described in the data preprocessing section. Specifically, fixation duration, saccade characteristics, and pupil size were extracted. Fixation duration was used as an indicator of visual focus \cite{Liu22, Chen11, Chien15}, saccade measures captured eye-movement dynamics between fixations \cite{Goldberg99}, and pupil size served as a physiological indicator of cognitive activity. Together, these measures were used to assess attention allocation and cognitive effort during password creation and memorability assessment.

\subsubsection{Self-Reported Measures}

In the demographic survey, participants reported their age, gender, highest completed education level, and primary occupation. They also indicated whether they had educational or professional experience in computer science or related fields and reported whether they used generative AI tools. Participants further reported how often they check password-strength meters and the importance of memorability and security. In the experiment, they rated the memorability and strength of their chosen passwords on five-point Likert scales.

\subsection{Data Processing}
Password strength was assessed using Shannon entropy~\cite{Shannon48}, a widely adopted baseline measure applied to all participant- and model-generated passwords~\cite{Shay10,Taha13,Thai24,Grassi2017}. Gaze data were preprocessed prior to analysis. Monocular gaze points on the display area for both eyes were fused into binocular coordinates, and pupil diameters were averaged across eyes. Pupil diameters were smoothed using the Savitsky-Golay filter and baseline-corrected by subtraction.

Identification by Velocity-Threshold (I-VT) was used to determine saccades and fixations \cite{Kasneci2024, Salvucci0}. Each sample was classified as a fixation candidate (for angular eye velocity <30°), a saccade candidate (>30°), or an outlier \cite{Tobii}. Fixations were only accepted if their duration lay between 0.075–1.0 s; saccades were only accepted if between 0.025–0.07 s. To relate the gaze behavior to survey behavior, each gaze point was assigned to an Area of Interest (AOI) based on recorded coordinates and timestamps.


\section{Results}


\subsection{Model Effects and Performance}

Passwords generated by different sources showed apparent differences in memorability performance and objective strength. User-generated passwords showed the highest memorability, whereas passwords suggested by AI-based and rule-based models were selected correctly less often (see Table \ref{tab:recall_count_model}). Logistic regression confirmed some of these differences. All model-generated passwords had lower odds of successful memorability compared to user-generated passwords (see Table \ref{tab:recall_model_userBase}). All these results were significant. Other pairwise comparisons did not reach significance thresholds; however, the estimates were directionally higher for AI-generated passwords than for the random generator.

\begin{table}[h]
    \centering
    \footnotesize
    \caption{Memorability by models (counts, totals, and success rates).}
    \label{tab:recall_count_model}
    \begin{tabular}{lcclc}
        \toprule
        & \multicolumn{2}{c}{\textbf{Correct Selection}} & \textbf{Total} & \textbf{Success Rate (\%)} \\
        \cmidrule(lr){2-3}
        \textbf{Model} & \textbf{0} & \textbf{1} & & \\
        \midrule
        DeepSeek-API     & 5 & 5  & 10 & 50.00 \\
        GPT-API         & 4 & 6  & 10 & 60.00 \\
        PassGPT        & 6 & 10 & 16 & 62.50 \\
        Random Generator & 4 & 1  & 5  & 20.00 \\
        User Generated & 4 & 75 & 79 & 94.94 \\
        \bottomrule
    \end{tabular}
\end{table}

\begin{table}[h]
    \centering
    \footnotesize
    \caption{Logistic regression comparing memorability success across password models (reference: \textit{user-generated}). Coefficients represent log-odds; \textit{OR} = odds ratio.}
    \label{tab:recall_model_userBase}
    \begin{tabular}{lccccc}
        \toprule
        \textbf{Variable} & \textbf{Coef} & \textbf{StdErr} & \textbf{z} & \textbf{p} & \textbf{OR} \\
        \midrule
        Intercept & 2.931 & 0.513 & 5.712 & 0.0000 & 18.750 \\
        DeepSeek-API & -2.931 & 0.814 & -3.599 & 0.0003 & 0.053 \\
        GPT-API & -2.526 & 0.825 & -3.063 & 0.0022 & 0.080 \\
        PassGPT & -2.420 & 0.728 & -3.325 & 0.0008 & 0.089 \\
        Random Generator & -4.317 & 1.230 & -3.510 & 0.0004 & 0.013 \\
        \bottomrule
    \end{tabular}
\end{table}

\begin{figure*}[ht!]
    \centering
    \begin{subfigure}[t]{0.48\textwidth}
        \centering
        \includegraphics[width=\linewidth]{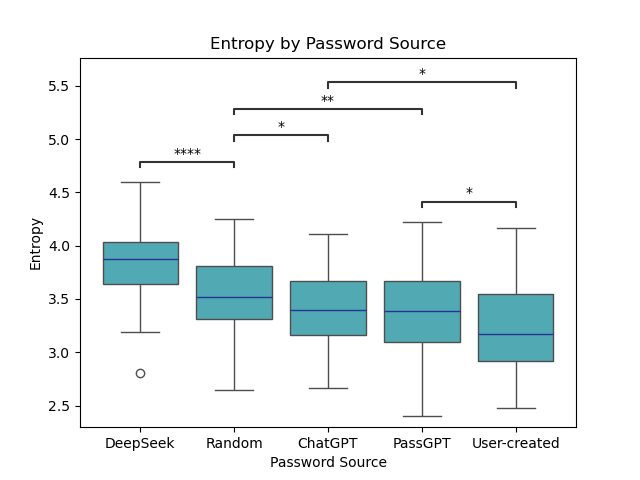}
        \caption{Entropy distributions by password source.}
        \label{fig:entropy_model_box}
    \end{subfigure}
    \hfill
    \begin{subfigure}[t]{0.48\textwidth}
        \centering
        \includegraphics[width=\linewidth]{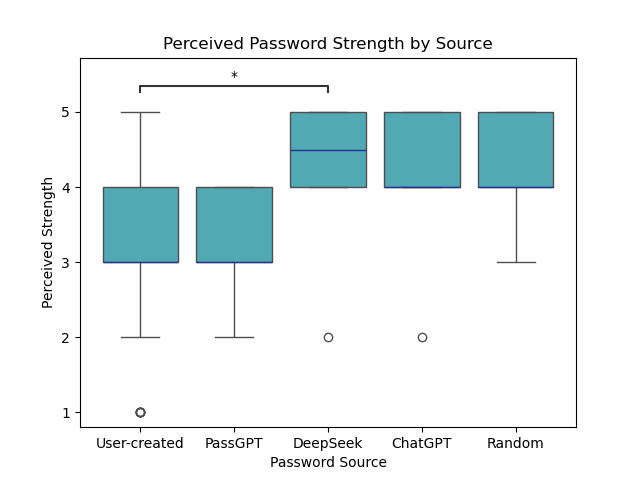}
        \caption{Perceived strength ratings by generation model.}
        \label{fig:strength_model_box}
    \end{subfigure}

    \vspace{0.1cm}

    \begin{subfigure}[t]{0.31\textwidth}
        \centering
        \includegraphics[width=\linewidth]{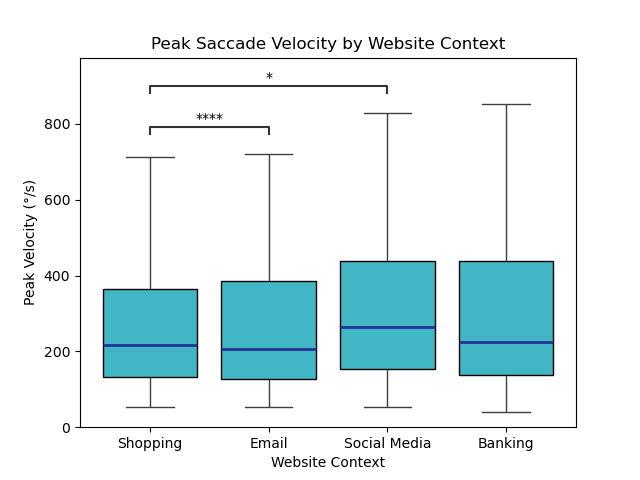}
        \caption{Peak saccade velocities by context.}
        \label{fig:vel_context}
    \end{subfigure}
    \hfill
    \begin{subfigure}[t]{0.31\textwidth}
        \centering
        \includegraphics[width=\linewidth]{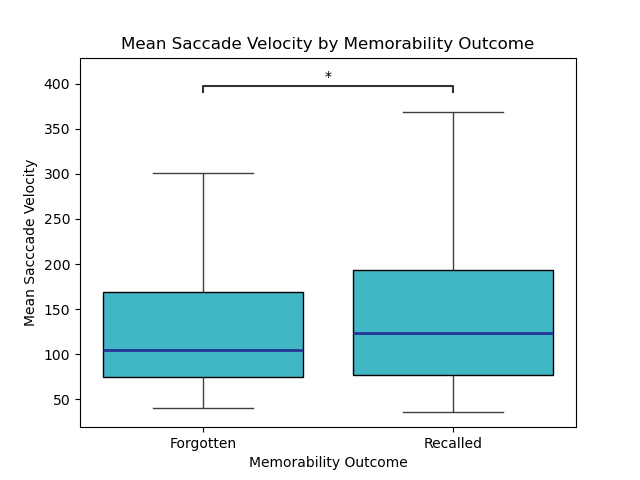}
        \caption{Mean saccade velocities vs. memorability.}
        \label{fig:mean_vels}
    \end{subfigure}
    \hfill
    \begin{subfigure}[t]{0.31\textwidth}
        \centering
        \includegraphics[width=\linewidth]{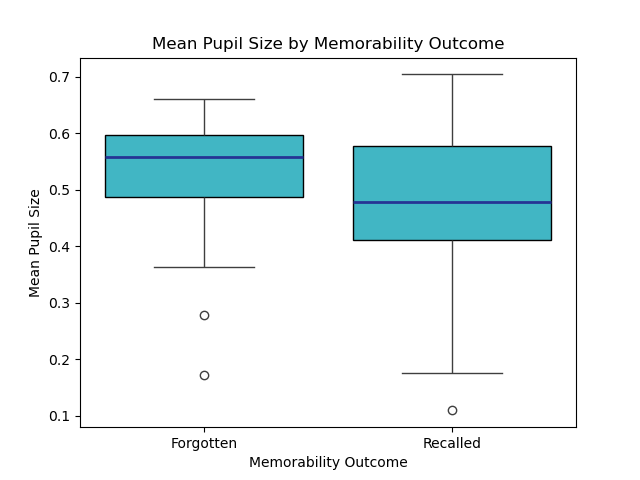}
        \caption{Pupil size vs. memorability outcome.}
        \label{fig:pupil_recall}
    \end{subfigure}
    \caption{Boxplot overview. Top row: objective entropy (a) and subjective strength ratings (b) across password models. Bottom row: gaze dynamics — saccade velocities by context (c), saccade velocity as memorability predictor (d, $p = .046$), and pupil size during recall (e, $p = .116$).}
    \Description{Five boxplots arranged in two rows. Top row: (a) entropy distributions across five password sources (DeepSeek, Random, ChatGPT, PassGPT, User-created) with significance brackets, showing AI-generated passwords have higher entropy; (b) perceived strength ratings on a 1-5 scale across the same sources. Bottom row: (c) peak saccade velocities across four website contexts (Shopping, Email, Social Media, Banking); (d) mean saccade velocities comparing Forgotten vs Recalled passwords; (e) mean pupil size comparing Forgotten vs Recalled passwords.}
    \label{fig:all_boxplots}
\end{figure*}

In contrast, entropy analysis revealed the opposite pattern. AI-generated passwords exhibited significantly higher entropy than user-generated passwords as shown in Figure \ref{fig:entropy_model_box}\footnotemark. Together, these results indicate a clear trade-off between strength and memorability: while AI-generated passwords tended to be objectively stronger, they were remembered less reliably than user-created ones. 

Visual attention did not differ across password generation models. Time spent viewing suggestions and gaze behavior were comparable across models, indicating that participants did not visually prioritize passwords based on their source. In contrast, subjective evaluations varied significantly by model. Perceived password strength differed across suggestion sources ($\chi^2$(16) = 33.99, \textit{p} = .005), with user generated passwords rated one step weaker than most other generators (with the exception of PassGPT). Figure \ref{fig:strength_model_box} illustrates the distribution of strength ratings across models.

\footnotetext{Significance annotations in boxplots follow the standard notation: * = $p \leq .05$, ** = $p \leq .01$, *** = $p \leq .001$, **** = $p \leq .0001$. The default significance threshold was $\alpha = .05$.}

\subsection{Context Effects}

Across the four website contexts: banking, email, shopping, and social media, password entropy and strength remained the same. The type of website or imagined usage scenario did not influence memorability or entropy of created passwords. Participants did not appear to adjust their passwords to the importance level implied by the context. 

Interaction between user behavior and website context revealed minimal context-driven variation. In saccade dynamics, we observed significantly higher peak and mean saccade velocities in the shopping context compared to email and social media (see Figure \ref{fig:vel_context}), indicating slightly more dynamic eye movements in this scenario.

\subsection{Attention to Context Drives Entropy}

While the explicit website context (e.g., Banking vs. Social Media) did not significantly alter password strength, our eye-tracking analysis revealed a significant positive correlation between visual attention to contextual cues and password quality. Specifically, a longer dwell time on the website logo during the input phase was positively correlated with password entropy (Spearman's $r = 0.26$, $p = .004$). 

We additionally controlled for input-phase duration. Logo engagement remained a significant predictor of entropy ($\beta = 0.082$, $p = .001$), suggesting that the observed relationship is not solely driven by longer interaction times.

This indicates that users who visually ground themselves in the service context by focusing on the logo may unconsciously prioritize the security requirements of that specific domain. In contrast, users who ignored visual context cues tended to default to weaker, context-agnostic passwords. This finding suggests that visual engagement serves as a stronger behavioral prime for security than the semantic category of the service alone.

\subsection{Gaze Dynamics and Memorability Performance}

Behavioral indicators provided insights into participants' cognitive state  during the memorability assessment. We identified a significant relationship between eye-movement dynamics and memory retrieval. Mean saccade velocity was a significant predictor of memorability, with higher mean saccade velocities associated with increased odds of successful memorability ($p = .046$) as seen in Figure \ref{fig:mean_vels}. This suggests a mechanistic difference in visual search strategies: users who correctly selected their passwords exhibited faster, more confident scanning behavior (recognition strategy), whereas users who failed to select correctly exhibited slower, hesitant eye movements (search strategy). Pupil size was not significantly associated with memorability outcomes in this sample (See Figure \ref{fig:pupil_recall} $p = .116$).

\section{Discussion}


\subsection{The Illusion of Security in AI Models}

The models’ almost opposite memorability and strength orders illustrate the security–usability trade-off. AI-based models generated passwords with higher entropy, yet these passwords had lower memorability. In contrast user-generated passwords showed the highest memorability and the lowest entropy. Despite this, participants continued to overestimate the quality of their self-generated passwords, failing to align their perceptions with the objectively lower strength. Although participants distinguished between password sources, perceived strength differences were markedly smaller than the corresponding entropy differences. This \textit{illusion of security} suggests that users judge strength based on familiarity rather than objective complexity.

\subsection{Visual Attention as a Security Prime}

Our most novel finding is the link between visual attention to contextual cues (logos) and password entropy. The modest but statistically significant correlation ($r=0.26$, $p=.004$) suggests that visual attention acts as a \textit{cognitive prime}. When users visually ground themselves in the service's context (by dwelling on the logo), they appear to subconsciously activate a security mindset appropriate to that context. In contrast, users who bypass these visual cues treat the task generically, leading to weaker passwords. This implies that security behavior is not just a product of knowledge, but of immediate visual engagement.

Furthermore, the significant relationship between saccade velocity and memorability supports eye tracking as a potential real-time sensor for authentication systems. High-velocity saccades indicate "recognition confidence," while low-velocity scanning signals indicate "search uncertainty," a distinction that could, in theory, be used to detect fraudulent users or impostors who are reading a stolen password rather than recalling it from memory.

\subsection{Design Implications: Visual Anchoring}

Current password meter designs focus on "after-the-fact" feedback (telling users their password is weak \textit{after} they type it). Our findings support a proactive design strategy we term \textbf{Visual Anchoring}. Since looking at the logo correlates with stronger passwords, UI designers should experiment with forcing attention to the service identity before the password field becomes active. For example, a login screen could briefly animate the service logo or require a micro-interaction with the brand image. This would cognitively "anchor" the user in the specific security context (e.g., "I am at my Bank") rather than the generic mechanical context (e.g., "I am typing a string"), potentially nudging them toward higher-entropy choices without explicit policy enforcement.

\paragraph{Privacy and ethical considerations.}
Real-world password workflows involve highly sensitive data; therefore, deployments should avoid external API calls and keep password processing local or within trusted environments to reduce leakage and profiling risks. Additionally, when gaze data are used, they should be treated as biometric data \cite{ozdel2024privacy} and handled with data minimization and strict access controls.

\subsection{Limitations}

The study included 15 participants; while sample size is not sufficient for broad population generalization, it allowed the observation of indicative oculomotor patterns related to attention and saccade dynamics. The use of fictional websites in a laboratory setting may have reduced perceived risk, potentially dampening the magnitude of context effects. Nevertheless, the robust statistical significance of the logo-dwell correlation ($p=.004$) within this sample suggests a meaningful underlying cognitive mechanic that warrants further investigation in future large-scale behavioral studies. 

\section{Conclusion}

This work examined password creation, memorability, and perceived quality across user-generated and AI-generated passwords using behavioral and eye-tracking measures. Results revealed a clear trade-off between security and usability. AI-generated passwords achieved higher objective strength but were less memorable, whereas user-generated passwords were more memorable despite lower entropy. Gaze measures provided insight into cognitive effort during password interaction, while subjective strength evaluation showed self-generation bias. Together, these findings highlight the importance of human-centered design in AI-assisted password systems. Most importantly, we identify Visual Anchoring as a promising design direction: guiding visual attention to service identity markers can unconsciously prime users for better security behaviors. Future work should validate this mechanism in larger-scale ecological settings.

\begin{acks}
The authors sincerely acknowledge Efe Bozkir for his substantial contributions to this work through extensive discussions and valuable feedback at multiple stages. We thank him for his continued mentorship and for the time and support he offered throughout this work.
\end{acks}

\bibliographystyle{ACM-Reference-Format}
\bibliography{references}
\appendix
\newpage
\section{GPT-4o / DeepSeek Password Generation Prompt}
\label{app:prompt_gpt}

\begin{figure}[h]
\begin{footnotesize}
\begin{verbatim}
You are Password Booster.
Receive a password as input and
immediately return a single, stronger
password derived from it.
Make it stronger but still memorable.
Your output MUST be exactly one line
and contain ONLY the new password.
No labels, no explanations, no
formatting, no markdown, no prefixes,
no suffixes. Just the password.
\end{verbatim}
\end{footnotesize}
\Description{The system prompt used for GPT-4o and DeepSeek password generation, instructing the model to act as a Password Booster that receives a password and returns a single stronger but memorable password with no additional formatting or explanation.}
\end{figure}

\subsection{Password Length Constraints Across Generators}

The password generation methods differed slightly in their length constraints:

\begin{itemize}
    \item \textbf{ChatGPT and DeepSeek:} No explicit length constraints were imposed. The models generated passwords freely based on the provided prompts.
    
    \item \textbf{PassGPT:} Passwords were generated with a minimum length of five characters longer than the user’s initial input, with an upper bound of 16 characters.
    
    \item \textbf{Random generator:} Passwords were constructed by appending a random number of additional characters (between 2 and 4) to the user’s initial password.
\end{itemize}

These differences should be considered when interpreting variations in password length and entropy across generation methods.

\begin{table}[h]
\centering
\caption{Password length statistics by generation method.}
\begin{tabular}{lccccc}
\toprule
\textbf{Source} & \textbf{Mean} & \textbf{SD} & \textbf{Min} & \textbf{Max} & \textbf{N} \\
\midrule
ChatGPT        & 13.57 & 3.26 & 8  & 24 & 120 \\
DeepSeek       & 19.08 & 4.12 & 12 & 30 & 120 \\
PassGPT        & 14.64 & 3.58 & 8  & 29 & 120 \\
Random         & 15.18 & 3.66 & 10 & 32 & 120 \\
User-created   & 12.15 & 3.41 & 8  & 28 & 203 \\
\bottomrule
\end{tabular}
\label{tab:pwd_length}
\end{table}

\end{document}